# Defect-Free Axially-Stacked GaAs/GaAsP Nanowire Quantum Dots with Strong Carrier Confinement

*Yunyan Zhang,[1,6]\* Anton V. Velichko,[2] H. Aruni Fonseka,[3] Patrick Parkinson,[4] James A. Gott,[3] George Davis,[2] Martin Aagesen,[5] Ana M. Sanchez,[3] David Mowbray[2] and Huiyun Liu[1]*

\* Correspondence: yunyang.zhang.11@ucl.ac.uk (Y.Z.)

1. Department of Electronic and Electrical Engineering, University College London, London WC1E 7JE, United Kingdom;

2. Department of Physics and Astronomy, University of Sheffield, Sheffield S3 7RH, United Kingdom

3. Department of Physics, University of Warwick, Coventry CV4 7AL, United Kingdom

4. School Department of Physics and Astronomy and the Photon Science Institute, University of Manchester, M13 9PL, United Kingdom

5. Center for Quantum Devices, Niels Bohr Institute, University of Copenhagen, Universitetsparken 5, 2100 Copenhagen, Denmark

6. Department of Physics, Universität Paderborn, Warburger Straße 100, 33098, Paderborn, Germany

**SUMMARY:**

Axially-stacked quantum dots (QDs) in nanowires (NWs) have important applications in fabricating nanoscale quantum devices and lasers. Although their performances are very sensitive to crystal quality and structures, there is relatively little study on defect-free growth with Au-free mode and structure optimisation for achiving high performances. Here, we report a detailed study of the first self-catalyzed defect-free axially-stacked deep NWQDs. High structural quality is maintained when 50 GaAs QDs are placed in a single GaAsP NW. The QDs have very sharp interfaces (1.8~3.6 nm) and can be closely stacked with very similar structural properties. They exhibit the deepest carrier confinement (~90 meV) and largest exciton-biexciton splitting (~11 meV) among non-nitride III-V NWQDs, and can maintain good optical properties after being stored in ambient atmosphere for over 6 months due to excellent stability. Our study sets a solid foundation to build high-performance axially-stacked NWQD devices that are compatible with CMOS technologies.

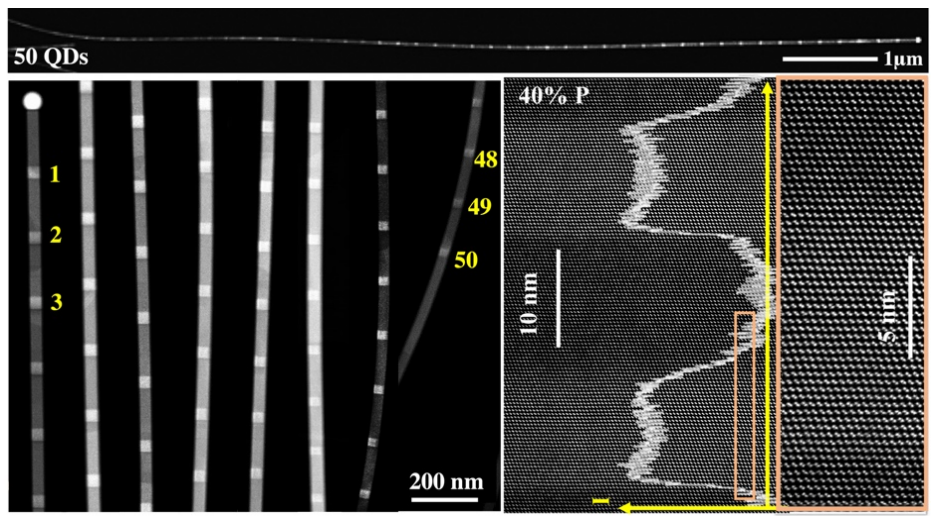

**INTRODUCTION**

Quantum dots (QDs) provide a unique semiconductor structure with fully quantised electronic states, permitting the fabrication of both high-efficiency classical and non-classical microelectronic and optoelectronic devices.[1,2] For example, nanosized lasers can be achieved by vertically stacking a large number (e.g. 50) of homogeneous QDs.[3] Enhanced complexity can be achived by closely stacking two or more QDs to achieve coupling/entangling between them electrically and optically.[4-6] These QD-molecular systems have been proposed as novel electromagnetic resonators, quantum gates for quantum computing, and thermoelectric energy harvestors *etc*.[7-10]

To date, the vast majority of QD physics and device studies have utilised self-assembled QDs, whose formation is strain driven via the Stranski-Krastanov growth mode.[11] However, the resulting strain field makes the growth of high-quality vertically-stacked identical QDs very challenging.[6] Horizontally-coupled QD pairs in thin-film structures are widely used in the study, but the center-to-center distance between two dots is large (>30 nm), resulting in significantly weaker coupling compared with the few nm separation of vertically stacked QDs.[9] Self-assembled QDs have a number of other significant disadvantages, including: formation at random positions, large inhomogeneous size distributions, limited shape and size control, and restrictions on which semiconductors can be combined in a single structure.

Semiconductor nanowires (NWs) have a unique one-dimensional morphology with many potential novel applications.[12-14] By changing the composition during NW growth, hetero-structures are formed, allowing the introduction of one or more quantum structure. Unlike self-assembled QDs, QD formation in a NW is not generally strain driven, permitting a greater range of semiconductor material combinations to be used. Direct control of both QD shape and size is possible, and the position of the QDs in a NW is fully controlled by the epitaxial growth parameters. Thus, identical QDs can be closely stacked, allowing the formation of molecular system.[15] or the gain region for lasers. Additional advantages include growth along the [111]B direction, which should minimise or eliminate exciton splitting, critical for the generation of entangled photons.[16,17] In addition, the small NW diameter provides high strain tolerance, which permits the combination of materials with large lattice and thermal expansion coefficient mismatches, allowing direct integration onto a Si platform.[18]

NW axial hetero-structures, including QDs, have been studied in a number of systems, e.g. InAsP-InP[19] and AlGaAs-GaAs[20] Achievements include very narrow spectral linewidths.[20-23], and the generation of both single photons and entangled photon pairs.[16,24,25] The majority of previous relevant studies have used NWs fabricated using the Au-catalyzed growth method,[26,27] possibly related to the easiness of structural control, including sharp QD interfaces and the crystal phases.[28-30] However, these NWs are incompatible with Si-based electronics, as Au can be incorporated into GaAs and InAs NWs at levels of the order of $10^{17}$–$10^{18}$ cm$^{-3}$.[31,32]

More recently, the non-foreign-metal-catalysed NW growth mode has been developed.[3] For example, GaAsSb-based multiple axial superlattices[33] and axially stacked InGaAs

QDs[3] have demonstrated reduced thresholds for single NW lasers. However, the majority of reports of QD growth still exhibit high-densities of stacking faults, with a mixture of zinc blende (ZB) and wurtzite (WZ) crystal structures.[34] A mixture of both crystal structures can have a significant impact on the electrical and optical properties.[35-38] Stacking fault formation is an especially serious issue for QDs grown by the widely-used self-catalyzed vapour-liquid-solid growth mode,[39] as the nano-sized group-III metal catalytic droplets are highly sensitive to the growth environment.[40] QD formation requires the switching of the growth flux to change the composition, thus producing a significant alteration of the growth environment, with the potential to introduce defects.[41] A big challenge, in growing closely stacked QDs, particularly with high P composition barriers, is to avoid changes in the growth environment which may lead to the generation of a high-density of defects.[34,41] To the best of our knowledge, there is only one report of the growth of defect-free stacked hetero-structure in self-catalyzed NWs , but with a diameter in the micrometre range;[42] full 3D confinement requires the NW diameters to be less than ~100 nm. The sensitivity of the catalytic droplets to their environment increases with reduced droplet size, due, for example, to the Gibbs−Thomson effect,[43], which makes it challenging to grow structures that are sufficiently small to exhibit true QD behaviour. Thus, there has been a lack of detailed studies of defect-free axially-stacked QD structures reproducibly grown by self-catalyzed methods.

In this work, we report an investigation of self-catalyzed defect-free GaAs/GaAsP single and multiple axially-stacked NWQDs with a deep carrier confinement. With robust surface passivation, the QDs exhibit very good optical properties, including narrow emission line widths well above liquid nitrogen temperatures, strong carrier confinement, and a large exciton-biexciton splitting, which are all beneficial for devices operating at high temperatures.

**RESULTS AND DISCUSSION**

**Defect-free GaAs/GaAsP QDs with various structures**

GaAsP NWs with phosphorous compositions of 20 or 40% and containing GaAs QDs of varying sizes were grown using a flux compensation technique (Supporting information S1). All NWs have a morphology similar to that shown in the scanning electron microscope (SEM) image of Figure 1A. It can be seen that the NWs have a highly uniform diameter of 50-60 nm along their entire length. For the single and two QD structures, the QDs are located around the NW mid-point (an example can be seen in the low magnification TEM image and higher magnification inset of the QD in Figure 1A). TEM analysis was carried out for QDs with different sizes. Figure 1B shows a ~10 nm thick single GaAs QD in a $GaAs_{0.8}P_{0.2}$ NW. The QD has a pure-ZB structure without the presence of any twins. Figure 1C shows the composition profile along the NW axis for a structure containing a taller, single QD than that is shown in Figure 1B. The QD is composed of almost pure GaAs (although EDX showed that some P was present in the QD, the amount was too low to be accurately quantified). To provide stronger carrier confinement, QDs with higher potential barriers are required. This is achieved by increasing the P content of the GaAsP NW. GaAs QDs with heights ranging from 5-30 nm were grown in GaAs$_{0.6}$P$_{0.4}$ NWs, all of which were found to exhibit a pure-ZB structure. Figs. 1D-F showing annular dark-field (ADF) images and respective average intensity profiles for ~5, ~10 and ~30 (image only) nm high QDs. The Matthews and Blakeslee limit for relaxation of a planar GaAs layer grown on GaAs$_{0.6}$P$_{0.4}$ is estimated to be ~15 nm, which is less than the current maximum QD height of 30 nm.[44] This suggests that the current structures are able to accommodate a relatively high degree of strain, a consequence of the small cross section of the NWs.

An important requirement for QDs is clearly defined boundaries with sharp interfaces. The interface abruptness should be maximised, especially, for the growth of small height QDs, otherwise the effective height of the QD is increased. The composition profiles shown in Figs. 1D and E demonstrate sharp QD interfaces. The lower (GaAsP-to-GaAs) interface is only ~6 monolayers ($\equiv$ 1.8 nm) wide and the upper (GaAs-to-GaAsP) interface is ~13 monolayers ($\equiv$ 3.6 nm). Both interfaces are significantly more abrupt than those observed in droplet-catalysed NWs where the group-III elements are switched (15-70 nm).[45,46] To the best of our knowledge, the current interface widths represent the smallest values for defect-free NW QDs where the group-V elements are switched. One reason for the sharp interfaces in the present structures is the low solubility of the group-V elements within the Ga droplet at the high growth temperature of 640 °C (Supporting Information 2).[30] The lower interface (growth of GaAs on GaAsP) is slightly sharper than the upper one due to the As/P exchange effect (Supporting Information 3). This is similar to the observation by Priante *et al.* in their heterojunctions grown in self-catalyzed AlGaAs NWs.[47] Moreover, their heterojunctions in Ref. 47 contain a high-density of stacking faults, which could be a result of uncompensated growth during source fluxes switching. In contrast, the interfaces in the current structures are defect-free.

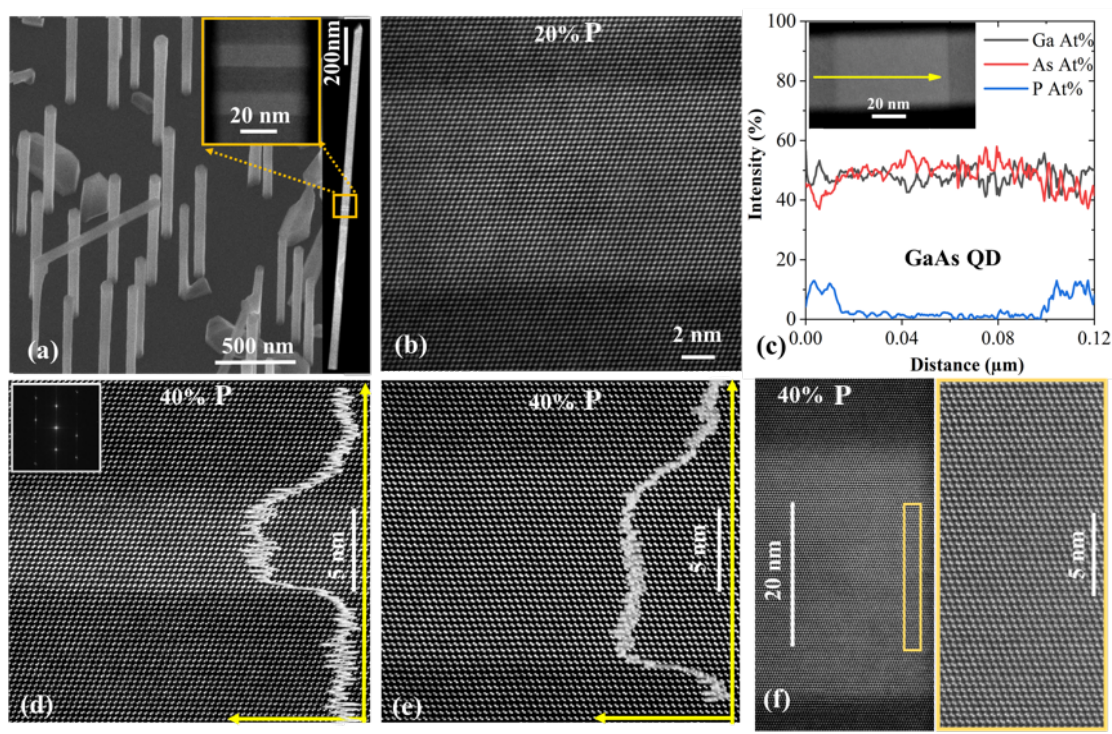

**Figure 1. Morphology and crystalline quality of GaAsP NWs with defect-free GaAs QDs.**
(**A**) 30°-tilted SEM image of GaAs$_{0.6}$P$_{0.4}$ NWs with two GaAs QDs located around the mid-point of the NW (see inset).
(**B**) High-magnification, ADF image of a ~10 nm-high GaAs QD within a GaAs$_{0.8}$P$_{0.2}$ NW.
(**C**) EDX composition profile along the axis of the GaAs$_{0.8}$P$_{0.2}$/GaAs QD shown in the inset. Annular ADF images of (**D**) ~5-nm-high, (**E**) ~10-nm-high and (**F**) ~30-nm-high GaAs QDs in GaAs$_{0.6}$P$_{0.4}$ NWs. The overlay curves in (D) and (E) are the integrated ADF intensity profiles. The inset in (D) is the SAED pattern for the region around the QD.

**Axially-stacked multi-QDs in a single NW**

To investigate if the QD separation affects their structural properties, QD pairs were

Page 8 of 31

grown with a small separation of ~10 nm. As seen in Figs. 2A and B these QD pairs are stacking-fault free, with very similar composition profiles and flat interfaces, indicating that the formation of closely spaced QDs with uniform properties is possible. This provides the potential to form QD molecules or large arrays of closely stacked QDs for laser applications.

To further investigate the suitability of the flux compensation method for the growth of axially-stacked QDs, a sequence of 50 nominally identical GaAs QDs were grown in a $GaAs_{0.6}P_{0.4}$ NW. These QDs appear as white segments in Figure 2C and D. The NW has a highly uniform diameter of 50 nm along its length and the round Ga droplet is retained at the tip, despite the long NW length of ~12 μm. Both of these observations indicate a highly stable growth environment during the long growth of 1.5 hours. 65% of the QDs are found to be defect-free (Figure 1E) and 31% contained a low density of twinning planes (1-4 / dot, as shown in Figure 2F). Only 4% of the QDs dots (2 dots) were found to have very thin WZ inserts, as shown in Figure 2G. This demonstrates the growth of a large number of axially-stacked QDs with a large percentage (96%) having high structural quality. The twinning typically occurs just above the GaAsP-to-GaAs (first) interface, indicating that the observed defects are formed at the beginning of the QD growth (red arrows in Figure 2F). This suggests a small droplet size fluctuation at the start of the QD growth due to the flux switching, which should be eliminated by further optimisation of the flux compensation technique. Neighbouring QDs have very similar separations, heights and composition profiles, as shown by composition scans along the NW axis (Figs. 2H and 2I).

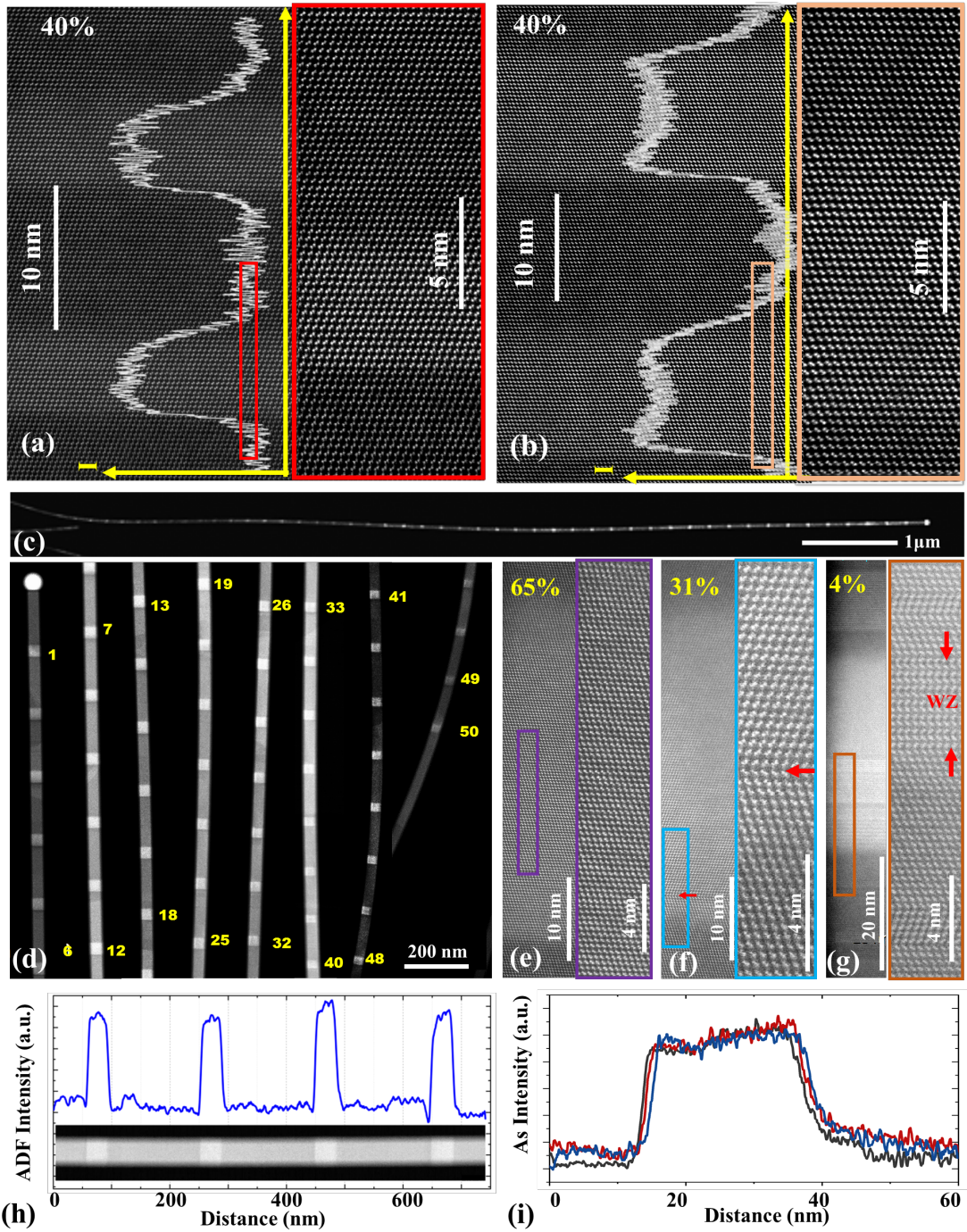

**Figure 2. Structural properties of axially-stacked GaAs$_{0.6}$P$_{0.4}$ / GaAs QDs.**

**(A)** and **(B)** Low-magnification (left) and high-magnification (right) ADF images of 5-nm-high and 10-nm-high closely-stacked pairs of QDs. The overlay curves are the integrated ADF intensity profiles.

**(C)** and **(D)** low-magnification ADF images of the entire NW containing 50 QDs.

High-magnification ADF images of representative QDs **(E),** without a defect , **(F)** with one twin plane, and **(G)** containing WZ segments , as indicated by the red arrows.

**(H)** axial integrated intensity profiles of the NW segment shown in the inset.

**(I)** As composition profiles for three adjacent QDs from the lower region of the NW.

**Long-term stability provided by insitu-grown passivation layer**

The emission properties of the QDs were studied by performing micro-photoluminescence (μ-PL) measurements on NWs that were transferred to a Si/SiO$_2$ substrate. The NWs had been stored in an ambient atmosphere for over 6 months, and the surface forms a thin oxidized layer, which can result in efficient non-radiative carrier recombination.[48] For GaAs/GaAs$_{0.6}$P$_{0.4}$ QDs with a bare surface (no passivation layer) only weak QD emission is observed, shown as the black spectrum in Figure 3. Following surface cleaning in a dilute ammonia solution (NH$_4$OH:H$_2$O=1:19) for one minute, stronger but very broad QD emission is observed, see blue spectrum in Figure 3. This indicates that the QD quality deteriorate severely with the exposure time in an ambient atmosphere and suggests that proper surface passivation is needed for achieving long-term stability. The small shift in the emission wavelength between these two spectra is in part due to inter-NW fluctuations, as it is not possible to study the same NW before and after cleaning. To improve the optical properties ~6 nm GaAs$_{0.6}$P$_{0.4}$ (to form 3D QD confinement), ~18 nm Al$_{0.5}$Ga$_{0.5}$As$_{0.6}$P$_{0.4}$, and ~9 nm GaAs$_{0.6}$P$_{0.4}$ shell layers were grown radially around the GaAsP core. These layers form a potential barrier to confine carriers within the core region and inhibit them from reaching the NW surface.[49] The sample with the additional passivation layers, exhibits emission consisting of only a single QD peak at low laser powers, with a narrow linewidth of ~500 μeV. This behaviour is preserved after over 6 months' storage in ambient atmosphere (the red spectrum in Figure 3), demonstrating the importance and effectiveness of the III-V passivation layers for long-term stability. This demonstrates that long-term stability of NW QDs can be achieved using only III-V materials fabricated during a single growth run, which can considerably simplify the passivation of NW QDs, and is very important for the widespread applications of NW QD devices. The QD emission for this sample is shifted with respect to the un-passivated sample as it is a significantly different structure with an enhanced 3D confinement. Unless otherwise stated, the measurements described below are for QDs passivated by this method.

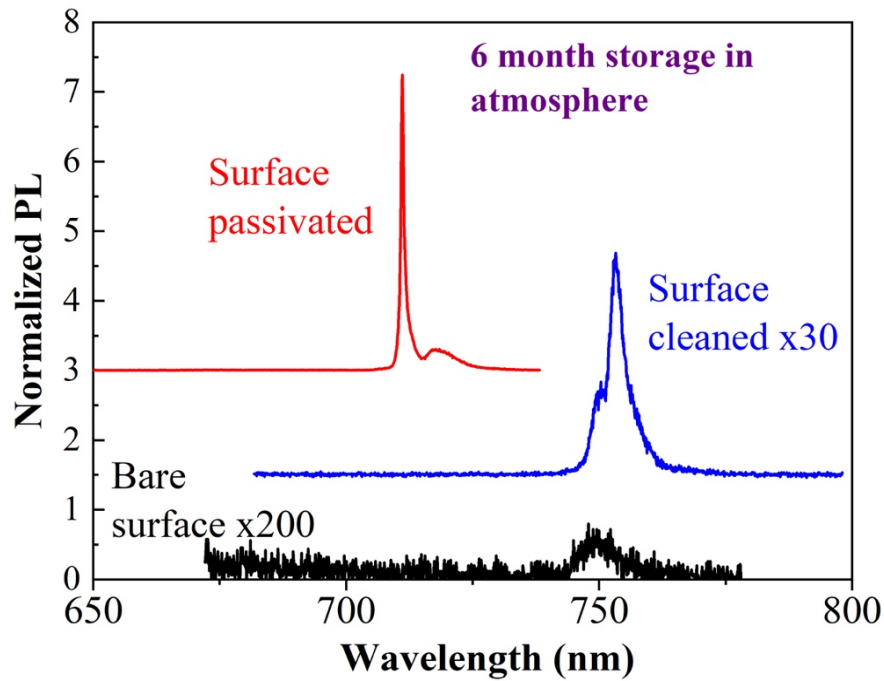

**Figure 3. Influence of insitu passivation on optical properties of NWQDs**. μ-PL spectra of GaAs QDs in GaAsP NW with an unpassivated surface (black), without surface passivation but with surface cleaning using an ammonia solution (blue), and with surface-passivation layers (red).

**Carrier dynamics in NWs with QDs**

The detailed optical emission properties of the QDs were studied by performing μ-PL measurements on surface passivated $GaAs_{0.6}P_{0.4}$ NWs with a core diameter of 50 nm and containing a single GaAs QD of nominal height 25 nm. The QD emission at 6K consists of a single peak at low laser powers (Figure 4A). Spectra (Figure 4A) recorded for different positions along the NW axis show that the emission is highly spatially localized, confirming it originates from the GaAs QD. The best fit to the measured profile is obtained by convolving a 1 μm width Gaussian function, representing the laser spot size, with a second Gaussian of width 0.7 μm. As this value (0.7 μm) is significantly larger than the physical size of the QD it represents the ability of photoexcited carriers to diffuse along the NW axis, followed by their capturing into the QD. Hence, the low temperature carrier diffusion length is ~0.35 μm. This indicates limited carrier diffusion at low temperatues, most likely a result of the slightly disorderd nature of the GaAsP alloy, and is in agreement with studies of the temperature behaviour of the QD emission intensity which are discussed below.

**Power-dependent behaviour of the NWQDs**

With increasing laser power, additional emission lines appear, as shown in Figure 4B. The inset plots the intensities of the two most spectrally resolved lines as a function of laser power. The different gradients are consistent with exciton and biexciton recombination and the lines demonstrate the expected high power saturation. Their separation is ~11 meV which is larger than previously reported values for non-nitride III-V QDs in a NW. For example, 6 meV for InAs QDs in GaAs NWs[50] and 3 meV for GaAsP QDs in GaP NWs[51] A large exciton-biexciton separation is due to the large confinement barriers (Supporting information S4), which is beneficial for single photon emission at elevated temperatures.

At higher laser powers, additional features appear above the energies of the exciton and biexciton lines in the μPL spectra of Figure 4B. These are attributed to higher order processes, either carriers in the ground state of the QD recombining in the presence of carriers in excited states, or direct recombination from the excited states. As the separations between confined QD states is comparable with the binding energies of exciton compexes, it is not possible to distinguish between these two mechanisms. Because of their strong spectral overlap, it is not possible to extract reliable power dependencies for these lines although over a limited power range the line at 689 nm demonstrates an exponent of 4.4, consistent with a higher order excitonic process.

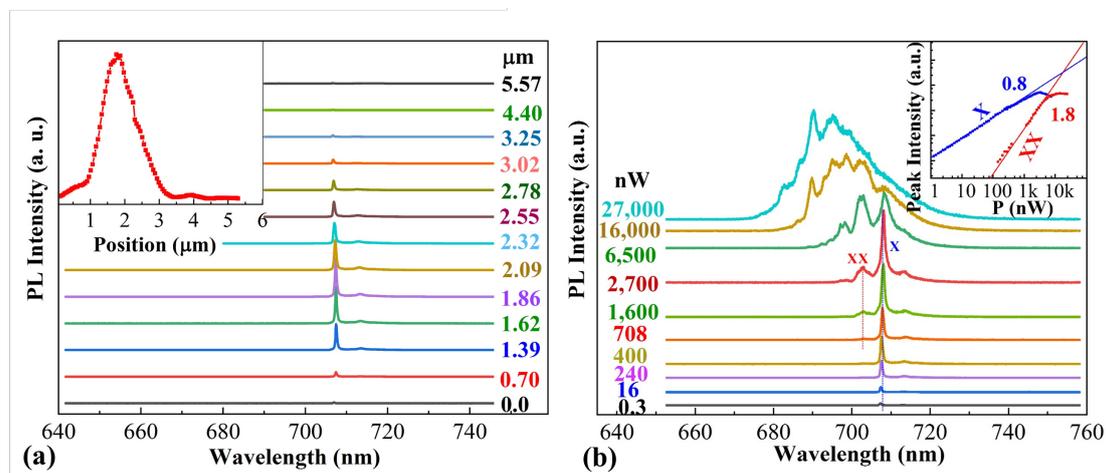

**Figure 4. Optical properties of a surface-passivated single ~25 nm GaAs dot in a ~50 nm diameter GaAs$_{0.6}$P$_{0.4}$ NW at 6K.**

(A) Position-dependent μ-PL spectra along the length of a NW. The laser power is 50 nW. The inset plots the intensity of the QD emission against exciting laser position.

(B) Power-dependent μ-PL spectra. The inset plots the intensities of two of the emission lines, X and XX, against laser power.

**Temperature-dependent behaviour of the NWQDs**

Figure 5A shows temperature dependent μ-PL spectra. With increasing temperature, the QD emission broadens; the full width at half maximum (FWHM) is plotted against temperature in Figure 5B. The solid blue line is a fit to the low temperature (≤140 K) data using the function

$$\Gamma(T) = \Gamma_0 + \frac{\Gamma_a}{\exp(E_a/kT) - 1}$$

where $\Gamma_0$ is the linewidth at low temperatures (1 meV for the current QD) and $\Gamma_a$ and $E_a$ are fitting parameters. This function describes broadening via the scattering of the excitons to a higher energy state by acoustic phonons, with $E_a$ being the energy separation of the two states[52] The function describes the experimental data well for temperatures up to ~140K (solid blue line in Figure 5B) and gives a value for $E_a$ of ~3 meV. Simulations performed using nextnano software[53,54] give confined electron and hole state separations for a 25 nm high and 40 nm diameter QD of ~11 and 6 meV, respectively. Hence, the determined value for $E_a$ is consistent with exciton scattering into an excited QD state. The linewidth at 140K is 9.8 meV. To the best of our knowledge, this is the first report of the emission linewidth at elevated temperatures for a non-nitride NWQD with long-term stability (Supporting information 5), which is in strong contrast to previous reports limited to temperatures below ~20K if the surface is not freshly cleaned. A linewidth of 9.8 meV is comparable to the lowest published values for nitride NWQDs at elevated temperatures (above ~100K).[55,56] By reducing the QD diameter, and hence increasing the electron and hole confined state separations, it should be possible to achieve smaller line widths at elevated temperatures. The large exciton and biexciton separation of 11 meV is larger than the emission linewidth of 9.8 meV at 140K, which allows the exciton and biexciton lines to be spectrally resolvable at this high temperature.

The temperature behaviour of the integrated emission intensity of a single QD exhibits a complex behaviour (Figure 5C). At low temperatures ($\lesssim$ 20K) a very small activation energy is found. Between 30-70K, the integrated intensity increases with increasing temperature. This can be explained by a change in carrier transport in the GaAsP barrier material. At low temperatures, carriers are relatively immobile due to localisation caused by alloy fluctuations. As the temperature increases, these carriers are thermally activated from the localisation centres and so a greater number are able to diffuse and be captured by the QD. Hence, there is a region where the QD emission intensity increases with increasing temperature. At high temperatures ($\gtrsim$ 80K) a large activation energy of ~90 meV is found. We have shown from studies of GaAs QWs in GaAsP NWs that both deep electron and hole confinement is achieved, a result of the mixed group-V structure and the large GaAs compressive strain.[57] Nextnano simulations indicate that the activation energy for electrons is 39% of the total energy separation of the barrier and QD bandgaps, calculated as 270 meV for the present structure. This suggests an electron activation energy of 105 meV, which is in reasonable agreement with the experimentally determined value of 90±5 meV.

Room temperature emission is observable from a GaAs QD grown in $GaAs_{0.6}P_{0.4}$ NWs and also QDs in NWs with lower P compositions. Figure 5D shows a sample with a $GaAs_{0.75}P_{0.25}$/GaAs NWQD and only a 30-nm thick GaAsP passivation layer. The origin of the main emission band is confirmed as arising from the QD by the spectral mapping shown in the inset of Figure 5D. High-temperature emission from QDs in a NW is typically observed for wide bandgap and high exciton binding energy materials (Supporting Information S5), e.g. GaN.[58,59] The observation of emission to relatively high temperatures further confirms the high crystalline, and hence, optical quality of the current QD structures.

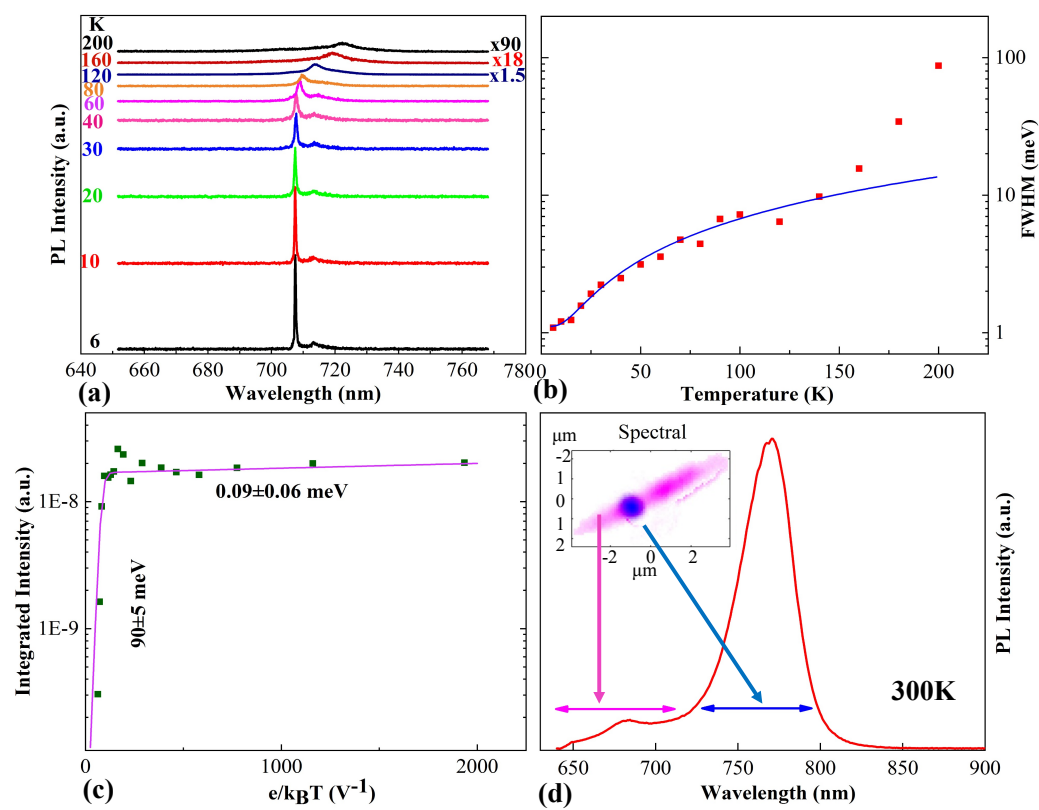

**Figure 5. QD temperature-dependent emission properties.**

**(A)** Temperature-dependent μPL spectra of a surface-passivated $GaAs_{0.6}P_{0.4}$/GaAs QD.

**(B)** QD emission linewidth and **(C)** integrated PL intensity plotted against temperature. The solid blue line in (b) is a fit to the low temperature data.

**(D)** PL spectrum at 300K created by combining spectra recorded separately from 640 individual

GaAs$_{0.75}$P$_{0.25}$ NWs each containing a single GaAs dot. The insets show a spectral map image of a representative NW, with two spectral bands used to create the image as indicated by the horizontal arrows in the main part of the figure.

**CONCLUSION**

In this work, we have demonstrated the growth of axially-stacked GaAs/GaAsP NWQDs with defect-free crystal quality and a deep carrier confinement. The QDs have interface widths as low as 6~13 monolayers, allowing the growth of QDs with small heights. It is also possible to form closely separated QD pairs with highly similar structural properties. The stacking of 50 GaAs QDs in one GaAsP NW is demonstrated, with 96% of the QDs exhibiting high crystalline quality. NWQDs degrade severely in atmosphere and surface passivation is shown to be critical for the optical properties of the QDs, particularly at high temperature. The addition of insitue (Al)GaAsP cladding layers results in a significant improvement in the optical properties and long-term stability when stored in ambient atmosphere. This demonstrates that long-term stability of NW QDs can be achieved using only III-V materials fabricated during a single growth run, which can considerably simplify the passivation of NW QDs, and is very important for the widespread applications of NW QD devices. The passivated QDs have a narrow linewidth of <10 meV at 140K; this is the first report of the high-temperature linewidth for a non-nitride III-V NWQD system after having been stored in ambient atmosphere for over 6 months, with the value comparable to the narrowest linewidths reported for nitride-based systems. The QDs exhibit a large carrier confinement energy of ~90 meV and emission up to 300K, consistent with a high crystalline quality, deep electron and hole confinement potentials and effective surface passivation. A large exciton-biexciton separation (~11 meV) is found. The narrow linewidth, emission at elevated temperatures and large exciton-biexciton separation are all requirements for high temperature operation of quantum emitters. Values for the current structures indicate the potential of non-nitride based NWQD quantum emitters to operate well above liquid nitrogen temperatures, which should greatly reduce device-operating costs and significantly increase the range of applications.

**EXPERIMENTAL PROCEDURES**

*NW growth*: The self-catalyzed GaAsP NWs were grown directly on Si(111) substrates by solid-source III−V molecular beam epitaxy. If not otherwise stated, the following growth parameters were used. The core $GaAs_{0.6}P_{0.4}$ ($GaAs_{0.8}P_{0.2}$) NWs were grown with a Ga beam equivalent pressure, V/III flux ratio, P/(As+P) flux ratio, and substrate temperature of $8.41\times10^{-8}$ Torr, ~30 (40), 41% (12%), and ~640°C, respectively. The GaAs QDs in $GaAs_{0.6}P_{0.4}$ ($GaAs_{0.8}P_{0.2}$) NWs were grown with a Ga beam equivalent pressure and V/III flux ratio of $8.41\times10^{-8}$ Torr and ~37 (44), respectively. To add passivation shell layers, the Ga droplets were consumed by closing the Ga flux and keeping the group-V fluxes open after the growth of the core. For samples used for optical measurement, GaAsP shells on $GaAs_{0.6}P_{0.4}$ ($GaAs_{0.8}P_{0.2}$) NWs were then grown with a Ga beam equivalent pressure, V/III flux ratio, P/(As+P) flux ratio, and substrate temperature of $8.41\times10^{-8}$ Torr, 110 (86), 49% (42%), and ~550 °C, respectively. ~18 nm $Al_{0.5}Ga_{0.5}As_{0.6}P_{0.4}$ (to block carriers from reaching the surface), and ~9 nm $GaAs_{0.6}P_{0.4}$ (to protect the AlGaAsP shell) shell layers were grown on the $GaAs_{0.6}P_{0.4}$ core. These layers form a potential barrier to confine carriers within the core region and inhibit them from reaching the NW surface. The $Al_{0.5}Ga_{0.5}As_{0.6}P_{0.4}$ shell was grown with an Al beam equivalent pressure, Ga beam equivalent pressure, V/III flux ratio, P/(As+P) flux ratio, and substrate temperature of $6.33 \times 10^{-8}$ Torr, $8.41 \times 10^{-8}$ Torr, 160, 49%, and ~550 °C, respectively. The substrate temperature was measured by a pyrometer.

***Scanning Electron Microscope (SEM)***: The NW morphology was measured with a Zeiss XB 1540 FIB/SEM system.

***Transmission electron microscopy (TEM)***: TEM specimens were prepared by simple mechanical transfer of the nanowires from the as-grown substrate to the holey carbon grid. The TEM measurements were performed with a JEOL 2100 and doubly−corrected ARM200F microscopes, both operating at 200 kV.

***Photoluminescence (PL)***: μ-PL spectra were obtained from single NWs, which had been removed from the original substrate and transferred to a new Si wafer. μPL spectra of single NWs were excited by a cw 515 nm diode laser. The samples were measured under vacuum inside a continuous flow cryostat (base temperature 6 K). The incident laser was focused with a 20x long working distance microscope objective to a spot size of ~1 μm diameter. The resultant PL was collected by the same microscope objective and focused into a 0.75 m spectrometer, where the spectral components were resolved and detected using a 300 l/mm grating and a nitrogen cooled Si CCD. The spectral resolution was ~0.5 meV. Higher resolution measurements were recorded using an 1800 lines/mm grating with a resolution of 0.09 meV.

Room temperature spectra were excited with 280 µW of 632.8 nm laser light focussed to a spot size of 0.8 µm diameter. Spectra of 898 individual NWs were recorded and the average of 640, which showed QD emission, used to create the spectrum of Figure 5D.

## ACKNOWLEDGEMENTS

The authors acknowledge the support of Leverhulme Trust, EPSRC (grant nos. EP/P000916/1, EP/P000886/1, EP/P006973/1), and the EPSRC National Epitaxy Facility. This project has also received funding from the European Union's Horizon 2020 research and innovation programme under the Marie Sklodowska-Curie grant agreement No 721394.

## AUTHOR CONTRIBUTIONS

YZ has written the manuscript and grown all the samples discussed under the supervision of HL. AVV and GD performed µPL measurements and optical data analysis directed by DM; AVV also performed nextnano simulations. HAF and JAG, performed the TEM and EDX measurements being directed by AMS. DM has performed calculations, general coordination and contributed to the manuscript writing. PP has performed room temperature µPL measurements and data analysis. MA, and GB contributed to manuscript discussion. DM, YZ, AVV, HAF and AMS all contributed to multiple revisions and finalising of the manuscript.

**DECLARATION OF COMPETING INTEREST**

The authors declare that they have no known competing financial interests or personal relationships that could have appeared to influence the work reported in this paper.

**SUPPLEMENTARY MATERIALS**

Supplementary information is available for this paper at https://doi.org/

Correspondence and requests for materials should be addressed to Y.Y.Z.

quantum dots. *Physical Review B*, *85*(8), 081303.

https://doi.org/10.1103/PhysRevB.85.081303

# Supporting Information

# Defect-Free Axially-Stacked GaAs/GaAsP Nanowire Quantum Dots with Strong Carrier Confinement

*Yunyan Zhang,[1,6]\* Anton V. Velichko,[2] H. Aruni Fonseka,[3] Patrick Parkinson,[4] James A. Gott,[3] George Davis,[2] Martin Aagesen,[5] Ana M. Sanchez,[3] David Mowbray[2] and Huiyun Liu[1]*

\* Correspondence: yunyang.zhang.11@ucl.ac.uk (Y.Z.)

1. Department of Electronic and Electrical Engineering, University College London, London WC1E 7JE, United Kingdom;

2. Department of Physics and Astronomy, University of Sheffield, Sheffield S3 7RH, United Kingdom

3. Department of Physics, University of Warwick, Coventry CV4 7AL, United Kingdom

4. School Department of Physics and Astronomy and the Photon Science Institute, University of Manchester, M13 9PL, United Kingdom

5. Center for Quantum Devices, Niels Bohr Institute, University of Copenhagen, Universitetsparken 5, 2100 Copenhagen, Denmark

6. Department of Physics, Universität Paderborn, Warburger Straße 100, 33098, Paderborn, Germany

**S1. Flux-compensation growth method**

To determine the factors leading to the formation of stacking faults, a comparison was made between two types of Ga-catalyzed GaAsP NWs that are formed during the growth of a single sample. These were formed within significantly different densities, and hence separation of both the NWs and clusters formed by parasitic growth between NWs. Type-I NWs are shown in Figure S1A. They exhibit a highly uniform diameter (50~60 nm) along their entire length

(3~4 μm), except very close to their base and tip where the diameter decreases. Type-II NWs are shown in Figure S1B and exhibit a noticeable tapering, with a gradually reduced diameter from the base towards the tip, indicating a decreasing droplet size during growth. Type-I NWs commonly have a pure ZB crystal structure except close to the tip and base. In contrast, the type-II NWs typically have a high density of stacking faults along their entire length. This suggests that the reduction of the droplet size correlates with the formation of stacking faults. It is found that the stacking faults at the base and tip of the NWs are generated by an unstable droplet at the start and end of the growth, when the source flux beams are switched on and off.

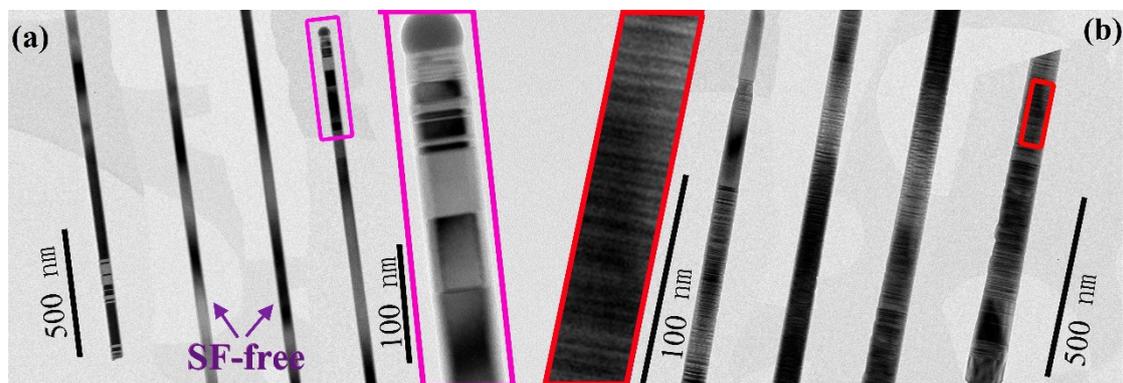

**Figure S1 | Relationship between droplet size fluctuation and the formation of stacking faults.** Transmission electron microscopy (TEM) images of GaAsP NWs with (**A**) highly uniform diameters and (**B**) a gradually reducing diameter from base to tip. The dark and sharp transverse lines are stacking faults.

To grow high quality axial hetero-structures with sharp interfaces, rapid switching of the growth fluxes is required. However, these flux switches have to be performed very carefully to avoid fluctuations of the droplet size, and hence the formation of stacking faults and the potential formation of WZ segments.[1-3] To avoid these fluctuations when growing axial hetero-structures, it is crucial to maintain the droplet super-saturation, which is achieved using a "flux compensation" method. When the P flux is turned off/on, the As flux has to increase/decrease accordingly, with the change in As flux volume proportional to that of the P flux volume. Our previous studies have shown that the P nucleation is stronger than that of As[4], so the compensating As flux should be larger than that of the P flux change. By varying the ratio (As flux change) : (P flux change) for the growth of a number of samples, with P compositions of either 20 or 40%, we have determined the optimum value to be between 1.48 and 1.80, with the precise value depending on the NW density, inter-NW parasitic growth density, growth temperature and III-V ratio. QDs grown outside this optimum compensation range tend to have a high-density of stacking faults. Figure S2 shows TEM images obtained from GaAs QDs grown in $GaAs_{0.8}P_{0.2}$ NWs using a compensation ratio of 2.2. As this compensation ratio is outside the optimum compensation range, the QDs of different heights (10~50 nm) contain a high-density of stacking faults; in contrast to the high quality, defect-free QDs grown within the optimum compensation range. This comparison clearly demonstrates the potential of the flux compensation technique and that its use is critical in obtaining high quality QDs.

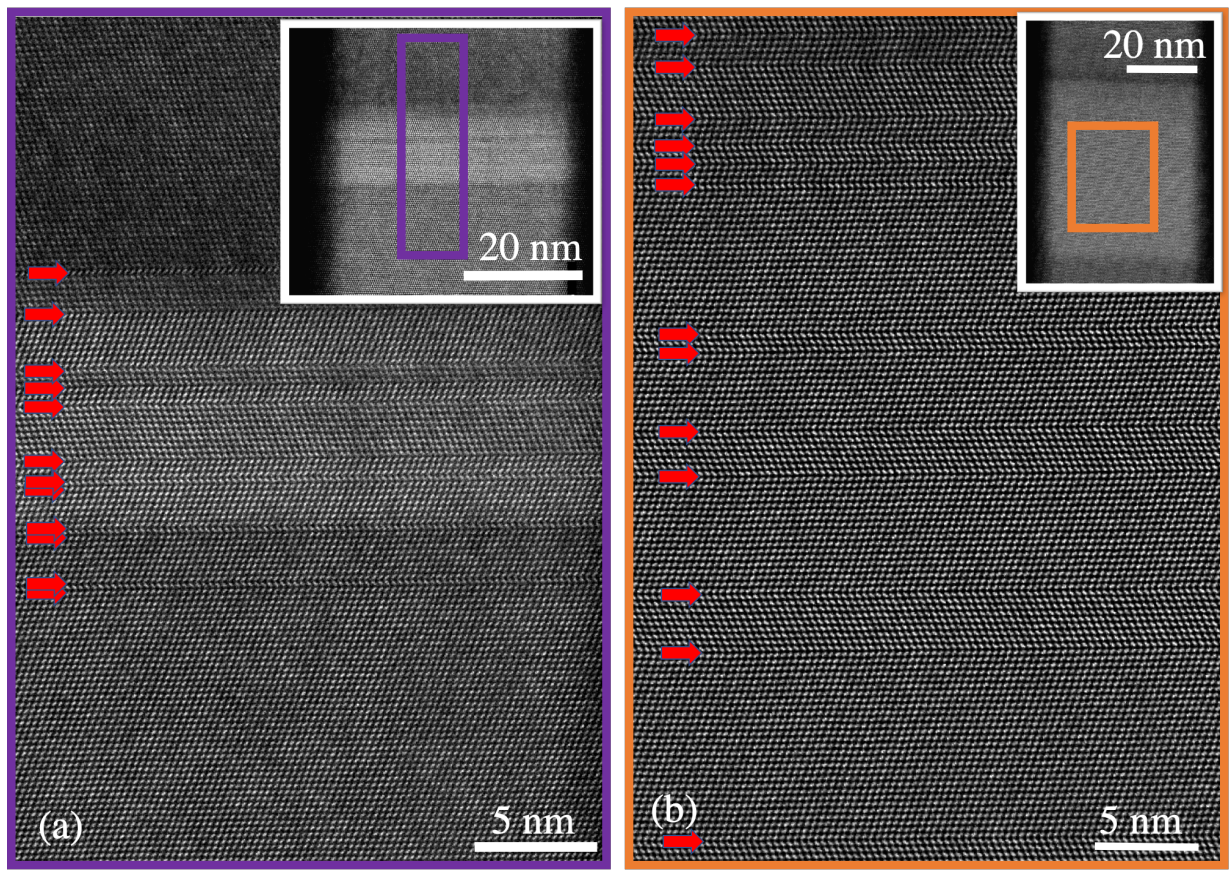

**Figure S2**. STEM images of GaAs QDs grown in $GaAs_{0.8}P_{0.2}$ NWs with a compensation ratio of 2.2. The QD height is ~10 nm in **(A)** and ~50 nm in **(B)**. The red arrows show the locations of twins. The insets are low-magnification TEM images of the NW segments containing the QDs.

## S2. Reduced reservoir effect for As and P in Ga metal droplets

The high solubility of group-III metals in the metal droplets used by the VLS method results in a significant reservoir effect; this can prevent the fabrication of sharp hetero-interfaces, especially interfaces that rely on a significant depletion of one element.[5] In contrast, group-V elements have a much lower solubility in the liquid metal compared to group-III metals, especially at high growth temperatures. This allows for very fast material depletion and switching.[6] The current GaAsP NW growth is performed at a relatively high temperature of 640 °C. This minimises the reservoir effect for As and P, resulting in the formation of sharp hetero-interfaces and the growth of almost pure GaAs QDs (Figure 1C, main paper).

## S3. Asymmetrical GaAs/GaAsP hetero interfaces

During compositional switching, As/P inter-diffusion occurs, reducing the sharpness of the interface. P atoms are more strongly bonded to Ga atoms, hence it is more difficult to replace P atoms with As atoms.[7] As a result, inter-diffusion at the GaAsP-to-GaAs interface is weaker than at the GaAs-to-GaAsP interface, leading to the former interface being sharper.

## S4. Sign of the biexciton binding energy

The biexciton emission occurs at a higher energy than that of the single exciton (Figure 4 of the main paper) indicating a negative biexciton binding energy. Although a negative binding energy is not possible in higher dimensionality structures, the complete confinement provided by a QD allows for the possibility of both types of biexciton binding energy (positive or negative). Previous reports of QDs in NWs have demonstrated both positive (e.g. In(Ga)As QDs in GaAs NWs[8]) and negative (e.g. InAsP QDs in InP NWs[9]) biexciton binding energies. The biexciton binding energy has been studied as a function of QD size in site-controlled pyramidal GaInAs/GaAs quantum dots [10] and self-assembled InAs QDs [11]. The sign of the biexciton binding energy will depend on the relative sizes of the Coulomb interactions between similar carriers (electron-electron and hole-hole) and different carriers (electron-hole). If the former dominates the latter, a negative binding energy will result. A mechanism via which this can occur is a significant physical separation between the electron and hole wave functions which enhances the Coulomb interaction between like carriers and reduces the interaction between different type carriers. In the site-controlled pyramidal GaInAs/GaAs QDs studied in Ref. *10*, a negative biexciton binding energy was observed for larger QDs and attributed to the effect of piezoelectric fields which separate the electrons and holes. Because of the [111] growth direction of the NWs and the strain applied to the GaAs QDs by the GaAsP barriers PZ fields are expected to be present in the current structures.

Figure S3 shows band edge profiles and the lowest electron and hole probability densities calculated using nextnano software and plotted along the central axis of the NW. The potential difference across the 25-nm high QD is ~80 meV, this gives an average PZ field of $3.2 \times 10^6$ $Vm^{-1}$. The combined effect of the strain modification to the band edge profiles and the strain-induced PZ field spatially separates the electron and hole wavefunctions along the NW axis, the maxima of the probability densities are separated by 9 nm. This separation is hence consistent with the observed negative biexciton binding energy observed in the current structure.

A significant physical separation of the electron and hole wave functions can enhance the exciton-biexciton energy difference; an additional contribution may arise due to the large confinement provided by the 40% composition GaAsP barriers. For example, it has been reported that InAs/AlAs QDs have a much larger exciton-biexciton splitting than InAs/GaAs QDs due to the deeper carrier confinement provided by the AlAs barriers.[12]

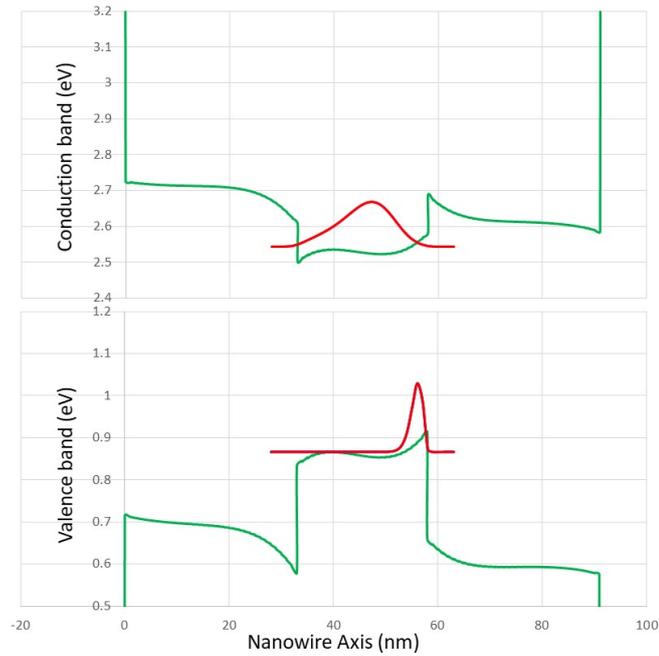

**Figure S3** Calculated conduction and valence band edge profiles along the central axis of the NW showing the effects of strain and strain-induced PZ fields. The red lines show the calculated probability densities for the lowest energy QD electron and hole states.

**S5. Summary of published NWQD emission linewidths as a function of temperature**

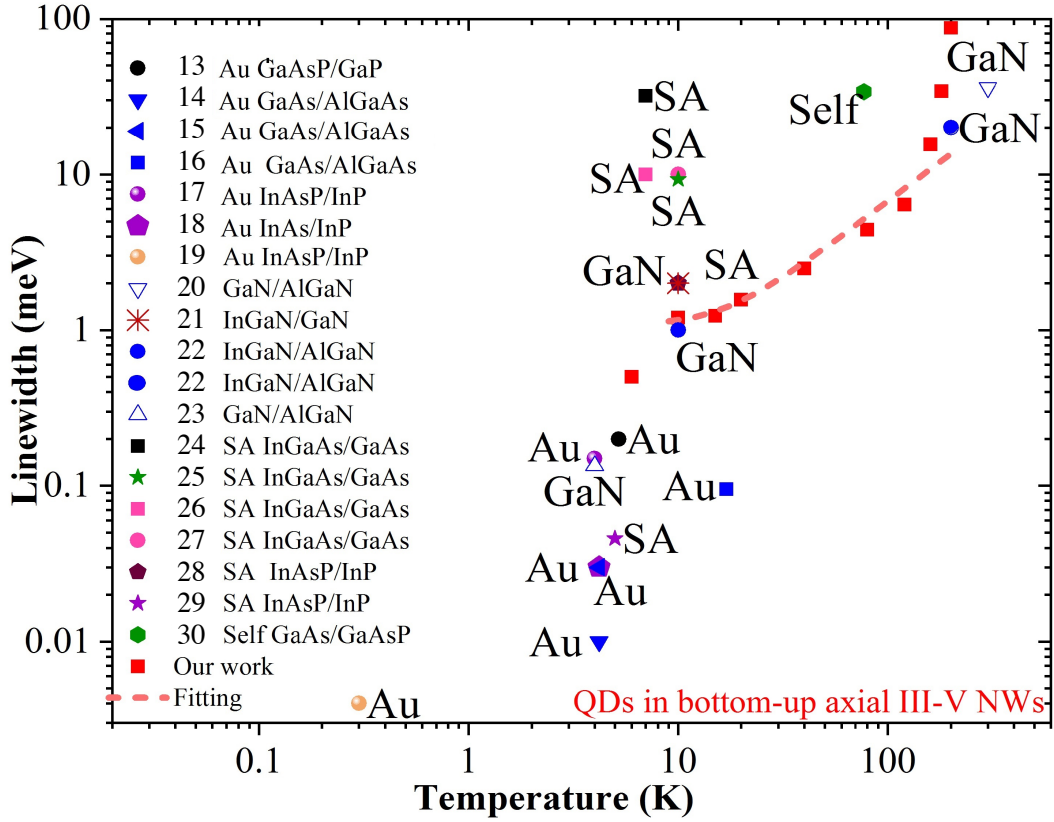

**Figure S4**. Emission linewidth summary for axial III-V QDs grown by bottom-up methods. The pink dash line is the fit to our temperature dependent data (Figure 5B). "SA" indicates selective-area growth. "Au" indicates Au catalysed growth and "GaN" indicates gallium nitride-based NWs.

**Table S1.** Emission linewidth summary for various III-V NWQD systems

| 1st Author | NW type | Material system | T (K) | Linewidth (meV) |
|---|---|---|---|---|
| **Bottom-up III-V QDs** | | | | |
| Borgstrom [13] | Au | GaAsP/GaP | 5.2 | 0.2 |
| Cirlin [14] | Au | GaAs/AlGaAs | 4.2 | 0.01 |
| Leandro [15] | Au | GaAs/AlGaAs | 4.2 | 0.03 |
| Heinrich [16] | Au | GaAs/AlGaAs | 17.0 | 0.095 |
| Haffouz [17] | Au | InAsP/InP | 4.0 | 0.15 |
| Dalacu [18] | Au | InAs/InP | 4.2 | 0.03 |
| Reimer [19] | Au | InAsP/InP | 0.3 | 0.004 |
| Holmes [20] | Nitride | GaN/AlGaN | 300.0 | 36.0 |
| Deshpande [21] | Nitride | InGaN/GaN | 10.0 | 2.0 |
| Deshpande [22] | Nitride | InGaN/AlGaN | 10 & 200 | 1-2 & 15-20 |
| Holmes [23] | Nitride | GaN/AlGaN | 4.0 | 0.135 |
| Tatebayashi [24] | SA | InGaAs/GaAs | 7.0 | 32.0 |

| Tatebayashi [25] | SA | InGaAs/GaAs | 10.0 | 9.3 |
| Tatebayashi [26] | SA | InGaAs/GaAs | 7.0 | 10.0 |
| Tatebayashi [27] | SA | InGaAs/GaAs | 10.0 | 10 |
| Yanase [28] | SA | InAsP/InP | 10.0 | 2.0 |
| Dorenbos [29] | SA | InAsP/InP | 5.0 | 0.046 |
| Wu [30] | Self-catalyzed | GaAs/GaAsP | 77.0 | 34 |
| **Our work** | **Self-catalyzed** | **GaAs/GaAsP** | **6.0** | **0.5** |
| **Our work** | **Self-catalyzed** | **GaAs/GaAsP** | **140.0** | **9.8** |
| **Other QDs in III-V NWs** | | | | |
| Bavinck [31] | Crysal Phase | InP | 4.2 | 0.023 |
| Kremer [32] | top-down | InGaAs/GaAs | 4.5 | 0.045 |
| Munsch [33] | top-down | InAs/GaAs | 4.2 | 0.004 |
| Francaviglia [34] | radial | GaAs/AlGaAs | 12.0 | 0.74 |
| Heiss [35] | radial | GaAs/AlGaAs | 4.2 | 0.029 |
| Shang [36] | radial | GaAs/AlAs | 4.2 | 0.037 |
| Weib [37] | radial | GaAs/AlGaAs | 5.0 | 1.0 |
| Yan [38] | radial | GaAs/InP | 5.5 | 1.28 |
| Yu [39] | radial | GaAs/AlGaAs | 4.2 | 0.1 |

Figure S4 and Table S1 summarize previous reports of emission linewidths for different III-V QDNW systems. The current work represents the first report of narrow emission linewidths for non-nitride based NWQDs above 20K. High-temperature emission from QDs in a NW is typically observed for systems with a wide bandgap and large exciton binding energy, e.g. GaN. Despite a much smaller band gap and exciton binding energy, we observe emission at 140K with a linewidth of 9.8 meV. This value is comparable with the best-reported values for nitride NWQDs.